\documentclass[mathleft,fleqn%
]{an}

\usepackage{graphicx}
\usepackage[varg]{txfonts}
\usepackage{amssymb}
\usepackage{graphicx}
\usepackage{natbib}
\usepackage[rightcaption]{sidecap}
\usepackage{txfonts}
\usepackage{units}
\usepackage{url}
\usepackage{pifont}
\usepackage[modulo, switch]{lineno}
\usepackage{hyperref}

\hypersetup{
    final=true,
    pageanchor=true,
    colorlinks=true,
    breaklinks=true,
    linkcolor=blue,
    citecolor=blue,
    urlcolor=blue,
    pdfpagemode=UseNone,
    pdftitle={A Method for Fitting Two-component He I 10830~\AA\ Profiles.},
    pdfauthor={S.J.\ Gonz{\'a}lez Manrique},
    pdfsubject={Astronomische Nachrichten},
    pdfkeywords={Sun: chromosphere, methods: data analysis, techniques: spectroscopic, line: profiles}}

\definecolor{blue}{rgb}{0.00, 0.00, 1.00}
\definecolor{red}{rgb}{0.86, 0.08, 0.24}
\definecolor{orange}{rgb}{1.00, 0.55, 0.00}
\definecolor{darkblue}{rgb}{0.00, 0.00, 0.55}
\definecolor{green}{rgb}{0.00, 0.39, 0.00}
\definecolor{pink}{rgb}{1.000000,0.078431,0.576471}

\begin{document}

% The following seven commands are intended for editorial usage and
% should be ignored by the author(s).
\Pagespan{1}{}% Document's page range. 
% If second parameter is left empty, the last page is computed
% automatically.
\Yearpublication{2014}%
\Yearsubmission{2014}%
\Month{0}%   
\Volume{999}%  
\Issue{0}% 
\DOI{asna.201400000}% 

\title{Fitting peculiar spectral profiles in \ion{He}{i} 10830~\AA\ absorption features}

\author{S.J.\ Gonz{\'a}lez Manrique\inst{1,2}\fnmsep\thanks{Corresponding author:
    \email{smanrique@aip.de}\newline}
    \and C.\ Kuckein\inst{1}
    \and A.\ Pastor Yabar\inst{3,4}
    \and M.\ Collados\inst{3}
    \and C.\ Denker\inst{1}
    \and C.E.\ Fischer\inst{5}
    \and P.\ G{\"o}m{\"o}ry\inst{6,1}
    \and A.\ Diercke\inst{1,2}
    \and N.\ Bello Gonz{\'a}lez\inst{5}
    \and R.\ Schlichenmaier\inst{5}
    \and H.\ Balthasar\inst{1}
    \and T.\ Berkefeld\inst{5}
    \and \\ A.\ Feller\inst{8}
    \and S.\ Hoch\inst{5}
    \and A.\ Hofmann\inst{1}
    \and F.\ Kneer\inst{9}
    \and A.\ Lagg\inst{8}
    \and H.\ Nicklas\inst{9}
    \and D. Orozco Su{\'a}rez\inst{3,4}
    \and D.\ Schmidt\inst{10}
    \and \\ W.\ Schmidt\inst{5}
    \and M.\ Sigwarth\inst{5}
    \and M.\ Sobotka\inst{7}
    \and S.K.\ Solanki\inst{8, 11}
    \and D.\ Soltau\inst{5}
    \and J.\ Staude\inst{1}
    \and K.G.\ Strassmeier\inst{1}
    \and M.\ Verma\inst{1}
    \and R.\ Volkmer\inst{5}
    \and O.\ von der L{\"u}he\inst{5}
    \and T.\ Waldmann\inst{5}}

\titlerunning{Fitting peculiar spectral profiles in \ion{He}{i} 10830~\AA\ absorption features}

\authorrunning{S.J.\ Gonz{\'a}lez Manrique et al.}

\institute{$^1$ Leibniz-Institut f{\"u}r Astrophysik Potsdam (AIP),
    An der Sternwarte 16, 
    14482 Potsdam, Germany\\
    $^2$ Universit{\"a}t Potsdam, Institut f{\"u}r Physik and
	Astronomie, Karl-Liebknecht-Stra\ss{}e 24/25,
	14476 Potsdam-Golm, Germany\\
    $^3$ Instituto de Astrof{\'i}sica de Canarias,
	c/ V{\'i}a L{\'a}ctea s/n, La Laguna 38205, Spain\\
    $^4$ Dept. Astrofísica, Universidad de La Laguna, E-38205, 
        La Laguna, Tenerife, Spain\\
    $^5$ Kiepenheuer-Institut f{\"u}r Sonnenphysik,
	Sch{\"o}neckstr. 6, 79104 Freiburg, Germany\\
    $^6$ Astronomical Institute of the Slovak Academy of Sciences, 05960
	Tatransk\'{a} Lomnica, Slovakia\\
    $^7$ Astronomical Institute, Academy of Sciences of the Czech Republic,
    Fri\v{c}ova 298, 25165 Ond\v{r}ejov, Czech Republic\\
    $^8$ Max-Planck-Institut f{\"u}r Sonnensystemforschung,
        Justus-von-Liebig-Weg 3,
        37077 G{\"o}ttingen, Germany\\
    $^9$ Institut f\"ur Astrophysik, Georg-August-Universit\"at 
        G\"ottingen, Friedrich-Hund-Platz 1,
        37077 G\"ottingen, Germany\\
    $^{10}$ National Solar Observatory, 3010 Coronal Loop
        Sunspot, NM 88349, USA\\
    $^{11}$ School of Space Research, Kyung Hee University,Yongin,
        Gyeonggi-Do,446-701, Korea}

\received{26 Jan 2016}
\accepted{\today}
\publonline{later}


\abstract{The new generation of solar instruments provides better spectral, spatial, and temporal 
resolution for a better understanding of the physical processes that take place on the Sun. 
Multiple-component profiles are more commonly observed with these instruments. Particularly, the 
\ion{He}{i} 10830~\AA\ triplet presents such peculiar spectral profiles, which give information 
on the velocity and magnetic fine structure of the upper chromosphere. The purpose of this investigation 
is to describe a technique to efficiently fit the two blended components of the \ion{He}{i} 
10830~\AA\ triplet, which are commonly observed when two atmospheric components are located within the 
same resolution element. The observations used in this study were taken on 2015 April 17 with the very 
fast spectroscopic mode of the GREGOR Infrared Spectrograph (GRIS) attached to the 1.5-meter GREGOR 
solar telescope, located at the Observatorio del Teide, Tenerife, Spain. We apply a double-Lorentzian 
fitting technique using Levenberg-Marquardt least-squares minimization. This technique is very simple and 
much faster than inversion codes. Line-of-sight Doppler velocities can be inferred for a whole map of 
pixels within just a few minutes. Our results show sub- and supersonic downflow velocities of up to 
32\,km\,s$^{-1}$ for the fast component in the vicinity of footpoints of filamentary structures. 
The slow component presents velocities close to rest. }

\maketitle

%##############################################################################
%
%    INTRODUCTION
%
%##############################################################################

\section{Introduction}\label{SEC1}

The spectral window around the \ion{He}{i} 10830~\AA\ triplet has often been used to study 
the magnetic properties and dynamics in the solar chromosphere. The height of formation of this triplet 
lies in the upper chromosphere \citep{Avrett1994}. It comprises three transitions, which take place 
between the lower $2^{3}S_{1}$ level and the upper $2^{3}P_{2, 1, 0}$ level. However, two transitions are 
blended and therefore only two spectral lines are observed. Generally, the blended line 
at $\sim$10830.30~\AA\ is called ``red'' component, while the non-blended line at 10829.09~\AA\ is called 
the ``blue'' component. The wavelengths were taken from the National Institute of Standards and 
Technology (NIST).\footnote{http://www.nist.gov/}

Of special interest are the strongly redshifted \ion{He}{i} profiles, which have been reported by 
several authors \citep[e.g.,][]{penn1995,muglach1997,muglach1998}. Commonly observed are two 
atmospheric components located within the same resolution element \citep{Lagg2007}. These ``dual flows'' 
were reported for the first time by \citet{Schmidt2000b} and have an easyly recognizable spectral 
pattern with two or more peaks next to the red component of \ion{He}{i}. One of the peaks is often 
subsonic while the other reaches supersonic velocities. At the average formation temperature of the 
\ion{He}{i} triplet, velocities close to 10\,km\,s$^{-1}$ correspond to the sound speed \citep{Lagg2007}. 
From now on we will label the profiles, which clearly show dual flow components of the \ion{He}{i} 
triplet as the slow (the one close to rest) and fast (redshifted with respected to the slow one) 
components. 

Several studies reported on dual or multiple-peak \ion{He}{i} profiles in different  phenomena on 
the Sun, e.g., \citet{Teriaca2003} in two ribbon flares, \citet{Sasso2007,Sasso2011} during a C2.0 
flare, \citet{Lagg2007} during an emerging flux process in an active region, or 
 \citet{AznarCuadrado2005,AznarCuadrado2007} in active and quiet-Sun regions. All 
of them reported supersonic velocities of the fast component between 40 and 90\,km\,s$^{-1}$. 
\citet{Lagg2007} suggested that these dual flows are formed in extremely filamentary structures and that 
their origin might be related to the magnetic field. It is expected that these profiles provide 
information on two different heights in the atmosphere.

%---- Figure 1 -----------------------------------------------------------------
\begin{figure}[t]
\includegraphics[width=\columnwidth]{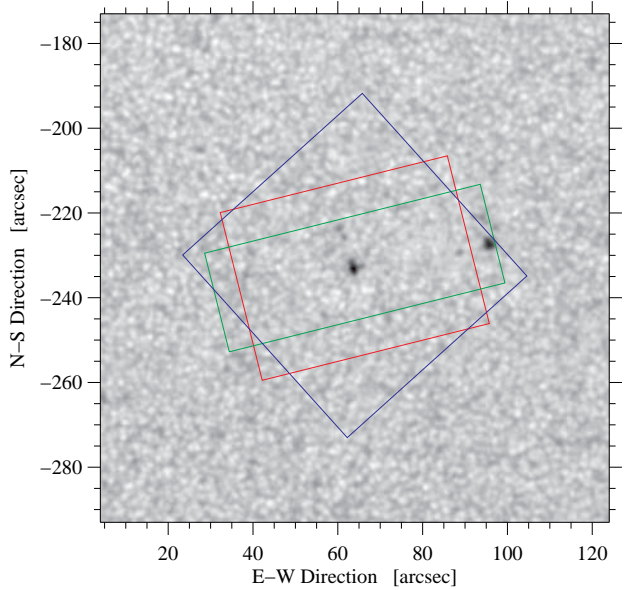}
\caption{Overlap of the FOV of the three instruments:
    GRIS (green), GFPI (red), 
    and CRISP (blue). The background continuum image is from SDO/HMI taken at 08:47~UT 
    on 2015 April 17.}
\label{FIG05}
\end{figure}
%-------------------------------------------------------------------------------

Several inversion codes are able to synthesize and invert the Stokes profiles of the \ion{He}{i} 
10830~\AA\ triplet, for example, the He-Line Information Extractor code 
\citep[H\footnotesize{E}\normalsize{LI}\footnotesize{X}\normalsize$^{+}$,][]{Lagg2004, Lagg2009}, the 
HAnle and ZEeman Light code \citep[HAZEL,][]{Asensio2008}, or the Milne-Eddington Line Analysis using a 
Numerical Inversion Engine \citep[MELANIE,][]{socas2001}. These codes infer the physical 
parameters such as, e.g., the line-of-sight (LOS) velocity and the vector magnetic field, among others. 
Usually these inversion codes require considerable computation time. In this paper, we 
present a simple and very fast technique to infer the LOS velocities for single and double-peaked 
profiles. Thus, the information present in the polarized Stokes parameters is disregarded. In 
particular, we adapted the technique to fit the slow and fast components of the \ion{He}{i} 10830~\AA\ 
triplet.

%---- Figure 2 -----------------------------------------------------------------
\begin{figure}[t]
\includegraphics[width=\columnwidth]{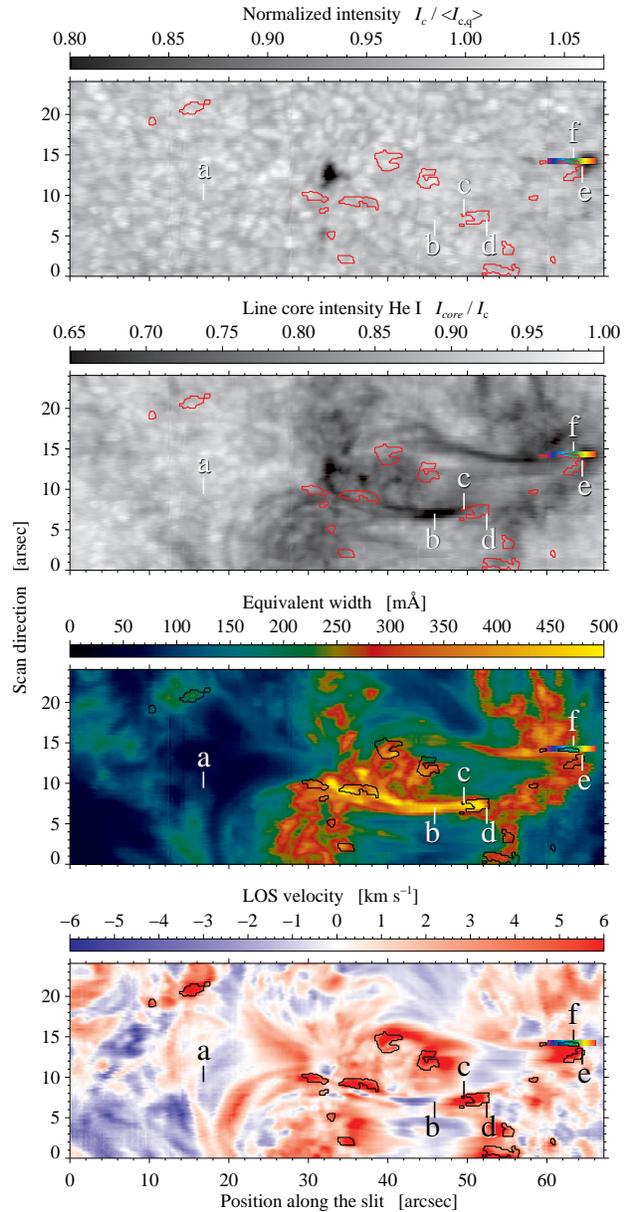}
\caption{GRIS slit-reconstructed images on 2015 April 17 at 08:42 UT of the observed region. From top to 
    bottom: continuum intensity, line core intensity of the red component of the \ion{He}{i} triplet, 
    equivalent width, and the one-component LOS velocity. The LOS velocity was inferred using a single 
    Lorentzian fit. The labels ``a'' to ``f'' mark the locations of six \ion{He}{i} profiles 
    shown in Fig.~\ref{FIG02}. The contours encompass only clearly discernible dual-flow components in 
    \ion{He}{i} profiles. The rainbow-colored bar in all maps marks the location of the profiles plotted 
    in Fig.~\ref{FIG04}.}
\label{FIG01}
\end{figure}
%-------------------------------------------------------------------------------

%##############################################################################
%
%    OBSERVATIONS
%
%##############################################################################

\section{Observations}\label{SEC2}

Our two-component fitting code was applied to one data set acquired with the 1.5-meter GREGOR solar 
telescope \citep{Schmidt2012, Denker2012} at Observatorio del Teide, Tenerife, Spain. An emerging 
flux region (EFR) located at heliographic coordinates S19 and W4 ($\mu \equiv \cos\theta = 0.97$) was 
observed on 2015 April 17 between 08:16~UT and 09:18~UT. Several instruments and telescopes were involved 
in this campaign. We will concentrate on the GREGOR Infrared Spectrograph \citep[GRIS,][]{Collados2012}.

For the first time the instrument was operated in the very fast spectroscopic mode (no polarimetry) in 
the \ion{He}{i} 10830~\AA\ spectral region. Typically, in polarimetric mode several 
integrations reduce the noise in the spectra but also prolongate the cadence of the scans. The fast mode 
relies on a single integration and accepts higher noise in exchange for an improved cadence. The 
standard deviation $\sigma = 3.3$\% of the normalized quiet-Sun continuum in the wavelength 
range 10835.8\,--\,10837.1~\AA\ for all spectra in the scan gives an estimate of the noise level, which  
will slightly increase for darker features like pores.

The spectral window of about 18~\AA\ comprised the photospheric \ion{Si}{i} 10827~\AA\ line, the 
chromospheric \ion{He}{i} 10830~\AA\ triplet, the photospheric \ion{Ca}{i} 10834 and 10839~\AA\ lines, 
and several telluric lines among other weaker spectral lines. The spectral sampling was 
$\delta_{\lambda} = 18.03$~m\AA\ pix$^{-1}$, and the number of spectral points was $N_{\lambda} = 1010$. 
The slit was by chance in parallel to two small pores belonging to the EFR (see Fig.~\ref{FIG05}). The 
pixel size along the slit was $\sim$0$\farcs$136, thus covering a length of 66\farcs3. For each position 
of the slit, the integration time was $t = 100$~ms with only one accumulation. This allowed for very 
fast spectroscopic scans covering an height of 24\farcs12 (with 180 steps and a step size of 
$\sim$0$\farcs$134) in only $\sim$58~s. During roughly one hour of observations (62 min), the GREGOR's 
altitude-azimuthal mount introduced an image rotation, which had to be corrected \citep{Volkmer2012}. 
The total rotation angle for the period of observation was 22.3$^\circ$. However, the image rotation 
for each map is small (never larger than 0.4$^\circ$) and therefore negligible. 

In total 65 scans were taken covering an area of $66\farcs3 \times 24\farcs12$. Nevertheless, only one 
map (08:42--08:43~UT) with abundant two-component profiles will be presented in this work (see 
Fig.~\ref{FIG01}). The data were taken under good seeing conditions with real-time corrections provided 
by the GREGOR Adaptative Optics System \citep[GAOS,][]{Berkefeld2012}. Simultaneous multi-wavelength 
observations were carried out with additional instruments including the GREGOR Fabry-P\'erot 
Interferometer \citep[GFPI,][]{Puschmann2012, Denker2010a} at the GREGOR telescope and the CRisp Imaging 
Spectro-Polarimeter \citep[CRISP,][]{Scharmer2008a} installed at the Swedish Solar Telescope 
\citep[SST,][]{Scharmer2003a} located at Observatorio Roque de los Muchachos, La Palma, Spain. An 
overview of all the involved FOVs is shown in Fig.~\ref{FIG05} after alignment with a continuum image 
of the Helioseismic and Magnetic Imager \citep[HMI,][]{Scherrer2012, Schou2012} on-board the Solar 
Dynamics Observatory \citep[SDO,][]{Pesnell2012}. The analysis of the full data set and an investigation 
of the dynamics related to the EFR are deferred to a forthcoming publication.

\subsection{Data reduction}\label{SEC2.1}

Dark and flat-field corrections were applied as part of the data reduction. The proper continuum of the 
GRIS spectra was derived by comparing the average quiet-Sun spectrum $I_\mathrm{QS}(\lambda)$ with that 
of a Fourier Transform Spectrometer \citep[FTS,][]{Neckel1984}, which was convolved with a Gaussian to 
take into account the degradation of the observed spectrum by straylight \citep[see][for a description 
of the procedure]{AllendePrieto2004}. The average quiet-Sun profile was computed in an area where the 
\ion{He}{i} absorption was almost absent. Common fringes were removed by multiplying every spectrum by 
the ratio between the equivalent quiet-Sun mean profile and the quiet-Sun mean profile $I^{\prime} / 
I_{\mathrm{QS}}$. Dust particles along the slit and other bright artifacts had to be removed from the 
data by using an additional flat field based on the science data assuming that the granulation was 
uniform and isotropic. Abnormal intensity peaks were found along some spectra due to the bad pixels on 
the detector chip. The peaks were detected by calculating the wavelength derivative of the spectrum of 
each pixel and searching for absolute values above a suitable threshold. Abnormal intensities were then 
replaced by an interpolated value between the previous and the next pixels.

The wavelength calibration was carried out using the two telluric lines located next to the 
\ion{He}{i} 10830~\AA\ triplet. An area of the quiet Sun was chosen to compute an average intensity 
profile. A direct comparison between the separation of the two telluric lines of the averaged intensity 
profile and the atlas profile of the Fourier Transform Spectrometer \citep{Neckel1984} provided the 
dispersion (18.03~m\AA\ pix$^{-1}$). The first telluric line at 10832.108~\AA\ was then used as the 
wavelength reference. In addition, the wavelength scale was corrected for solar orbital motion and 
rotation, Earth's rotation, and the solar gravitational redshift. (see appendices A and B in 
\citealt{Kuckein2012b}).

The fast component is blended by a telluric line for high LOS velocities. Therefore, the 
telluric line was removed from our spectra. To accomplish this task we averaged all spectral profiles 
from the map to compute a mean intensity profile. We assumed that the telluric line has a constant 
wavelength, i.e., it does not shift along the wavelength axis at any time. The telluric line was fitted 
in the spectral range of 10831.48--10832.56~\AA\ with a single Lorentzian profile. Hence, we have 
computed a synthetic telluric profile. The telluric line was then removed by dividing each spectrum by 
the synthetic profile. A smooth transition of the spectrum at the location where the telluric line 
started was assured by forcing the continuum in the line wings of the telluric line to be $I_\mathrm{c} = 
1$. Therefore, each spectrum was normalized to the local continuum before removing the telluric line.

%---- Figure 3 -----------------------------------------------------------------
\begin{figure}[t]
\includegraphics[width=\columnwidth]{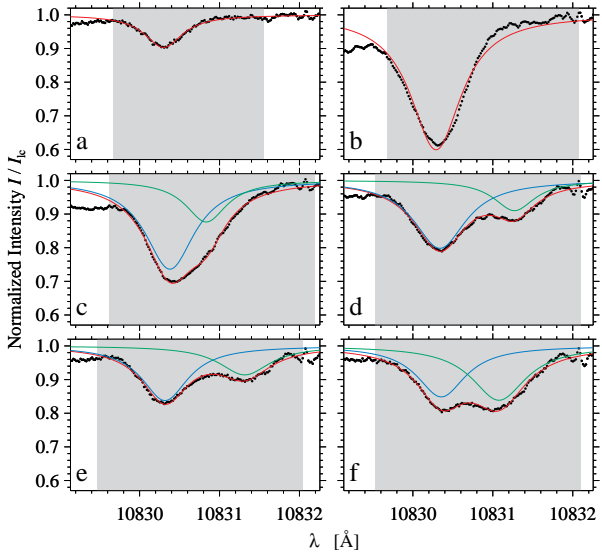}
\caption{The dots represent the observed \ion{He}{i} red-component profiles and the best 
    Lorentzian fit is shown with a solid red line. When dual-flow
    \ion{He}{i} profiles are present, the fit (solid red line) is a 
    superposition of two individual Lorentzians (blue and green solid lines). 
    Labels `a' to `f' show the locations of the profiles in the map of 
    Fig.~\ref{FIG01}. The grey background marks the fitting range.}
\label{FIG02}
\end{figure}
%-------------------------------------------------------------------------------

Figure~\ref{FIG01} shows a slit-reconstructed continuum intensity map (\textit{top}) and a line-core 
intensity map of the red \ion{He}{i} component (using the minimum of the line for each spectrum) 
normalized to the local continuum intensity $I_{core}/I_{lc}$ (\textit{second from top}). 

%##############################################################################
%
%    METHOD
%
%##############################################################################

\section{Method}\label{SEC3}
Most of the \ion{He}{i} 10830~\AA\ intensity profiles show the expected two spectral lines, which 
comprise the blue and red components. However, a few percent of these profiles, in our data set 
around 3\%, shows a clear signature of a fast component in the \ion{He}{i} 
triplet (see the contours in Fig.~\ref{FIG01}). As a first approach we assume that all profiles have 
only one component. Hence, we use a single Lorentzian profile to fit all \ion{He}{i} profiles:
%%%%%%%
\begin{eqnarray} \label{EQ01}
F & = & 1 - \frac{A_{0}}{u^{2} + 1} \quad \mathrm{with} \\
u & = & \frac{x - A_{1}}{A_{2}} ,
\end{eqnarray}
%%%%%%%
where A$_{0}$ is the amplitude, A$_{1}$ the peak centroid, A$_{2}$ the half-width-at-half-minimum 
(HWHM), and the equation is set to unity to normalize the synthetic intensity continuum. The profiles 
were fitted using the Levenberg-Marquardt least-squares minimization \citep{More1977, More1993} , where 
the MINPACK-1 software package was implemented in IDL by \citet{Markwardt2009}. An advantage of using 
this routine is that it is possible to impose an upper and lower bounding limit for each free 
parameter of Eq. \ref{EQ01} independently. As a first approach, we retrieved suitable limits by fitting 
the mean quiet-Sun  profile. The limits were then modified to increase the accuracy of the fits.
The fitted wavelength range is automatically adjusted depending on the amplitude of the spectral 
profile. The larger the amplitude of the line the broader the spectral range.
Since the dual-flow profiles are systematically redshifted we did not include the blue component 
in the analysis. 

The LOS velocities inferred with the single Lorentzian fits are shown in Fig.~\ref{FIG01} 
(\textit{bottom}). Positive velocities are related to downflows (red) while negative velocities (blue) 
represent upflows. Note that double-peaked profiles, i.e., dual-flow \ion{He}{i} profiles, 
are not well fitted with a single Lorentzian profile. Since the wavelength is given on an 
absolute scale, the wavelength reference for the LOS velocities was set to 10830.30~\AA, which 
corresponds to the average laboratory wavelength of the \ion{He}{i} red component (10830.25~\AA\ and 
10830.34~\AA\ respectively, from the NIST database). Figure~\ref{FIG01} also shows the equivalent width  
(\textit{second from bottom}) obtained from the single-Lorentzian fits. 

%---- Figure 4 -----------------------------------------------------------------
\begin{figure}[t]
\includegraphics[width=\columnwidth]{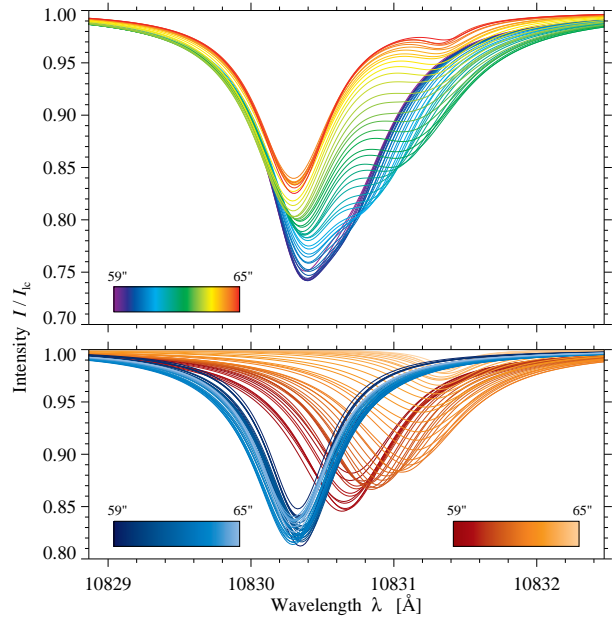}
\caption{Top: synthetic \ion{He}{i} dual-flow profiles using the fit parameters 
    retrieved from the double-Lorentzian fits. The color-coded bar extends over a 
    distance of $\sim$6\arcsec\ (see Fig.~\ref{FIG01}) linking profiles and positions. Note that 
    dual-flow components are hardly discernible in the first (violet) and last (red) profiles. 
    Bottom: decomposition of the dual-flow profiles into two Lorentzian (blue and red) components.}
\label{FIG04}
\end{figure}
%-------------------------------------------------------------------------------
 
The next step concerns fitting the dual-flow profiles with a double-Lorentzian profile:
%%%%%%%
\begin{eqnarray}
F & = & 1 - \frac{A_{0}}{u_{1}^{2} + 1} - \frac{A_{3}}{u_{2}^{2} + 1} \quad \mathrm{with} \\
u_{1} & = & \frac{x - A_{1}}{A_{2}} \quad\mathrm{and}\quad u_{2}  =  \frac{x - A_{4}}{A_{5}} ,
\end{eqnarray}
%%%%%%%
where A$_{0,3}$ are the amplitudes, A$_{1,4}$ are the centroids, and A$_{2,5}$ are the HWHM of each 
Lorentzian, respectively. Due to the very short computation time (a few minutes), all spectral profiles 
were fitted again with two Lorentzians. However, in the future we will adapt the code to only fit the 
two-component profiles. 

An adaptive wavelength range was not used this time. We chose two different wavelength ranges depending 
on the position of the line core of the \ion{He}{i} red component. This ignored the \ion{He}{i} blue 
component from the selected wavelength range. Therefore, if the line core of the \ion{He}{i} red 
component was located below $10830.49$~\AA, then the wavelength range was $[-0.83,+1.73]$~\AA\ with 
respect to the line core. Otherwise, if the position of the line core was $> 10830.49$~\AA, then the 
wavelength range was {$[-1.37,+1.73]$~\AA} with respect to the line core.

The initial estimates of the fit parameters $A_0$--$A_5$ were based on the single-Lorentzian fits and on 
the location of the deepest line core of the dual-flow components. The peak centroid of the fast 
component is always more redshifted than the slow component. The peak centroid of the slow 
component is always at rest. The amplitudes $A_0$ and $A_3$ are both initially set to the amplitude 
inferred from the single-Lorentzian fit, but are allowed to vary in the course of the fitting. The 
parameters $A_2$ and $A_5$ have a fixed value of the HWHM of 0.32~\AA\ (mean HWHM values of all the 
profiles fitted with a single Lorentzian).

In order to localize the dual-flow profiles, different types of these profiles were 
manually selected and then correlated with all the profiles of the map. The wavelength range for this 
correlation was between  [10829.50--10832.56]~\AA. In our particular case, we have 
selected more than 70 clearly distinguishable dual-flow profiles. Weak dual-flow profiles like, for 
instance, the first violet and red (too small amplitude) synthetic profiles shown in Fig.~\ref{FIG04}, 
were not selected because they are too weak to clearly identify the fast component. The median value of the 
correlation of all profiles were used as a threshold. Hence, only profiles which showed a larger 
correlation factor than  0.98\% were treated as two-component profiles. The contours in Fig.~\ref{FIG01} 
encompass these profiles. Developing an automatic algorithm to select the dual-flow profiles remains as 
future work.

%---- Figure 5 -----------------------------------------------------------------
\begin{figure}[t]
\includegraphics[width=\columnwidth]{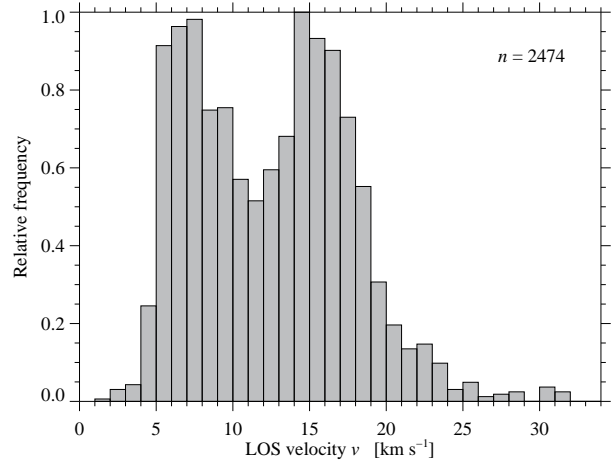}
\caption{Normalized frequency distribution of the LOS velocities of the 
    redshifted \ion{He}{i} component obtained with dual-flow fitting using 
    two Lorentzians. The variable $n$ is the total number of pixels which were used 
    for the histogram. The pixels
    belong to the areas encompassed by the contours shown in Fig.~\ref{FIG01}.}
\label{FIG03}
\end{figure}
%-------------------------------------------------------------------------------

%##############################################################################
%
%    Results
%
%##############################################################################

\section{Results}\label{SEC4}

Some examples to illustrate the quality of the fitting method are shown in Fig.~\ref{FIG02}. All profiles 
are also marked in Fig.~\ref{FIG01}. Profile `a' represents a common quiet-Sun profile with a LOS 
velocity close to zero ($-0.01$\,km\,s$^{-1}$). An example of a profile with a large equivalent width 
(472.4~m\AA) is shown in panel `b'. It arises from a dark filamentary structure in the chromosphere (see 
Fig.~\ref{FIG01}). However, the LOS velocity is rather small and only reaches $-0.3$~km\,s$^{-1}$ 
(upflow). Profiles `c'--`f' represent different types of dual-flow profiles. In particular, 
profile `c' was selected because of its deep line core, i.e., higher absorption. The LOS velocity of the 
fast component (solid green line) was about 14.8\,km\,s$^{-1}$ while the slow component (solid blue 
line) only showed about 2.2\,km\,s$^{-1}$. Profiles `d' and `e' are located near the footpoints of the 
filamentary structure. The latter had one of the largest inferred velocities of the fast component   
\mbox{($\sim$28.1\,km\,s$^{-1}$)}. The remaining profile `f' presents two components with similar 
amplitudes. The fast component reached a LOS velocity of about 21.5\,km\,s$^{-1}$. 

The normalized frequency distribution of the LOS velocities belonging to the \ion{He}{i} fast component 
is shown in Fig.~\ref{FIG03}. Velocities are in the range of 1--32\,km\,s$^{-1}$. The distribution is 
clearly double-peaked. The maxima are located close to 7\,km\,s$^{-1}$ (subsonic) and 16\,km\,s$^{-1}$ 
(supersonic), respectively. However, two distinct populations cannot be
identified in the FOV, either because the statistical sample is too small or
sub-/supersonic flows are transitory in a confined space. Our normalized frequency distribution 
is similar to the one reported by \citet{AznarCuadrado2007} in magnetic field-free areas. 

We traced the distribution of the dual-flow profiles along a given slit position. The slit position 
was chosen close to the footpoints of the filamentary structure and next to the pore. The rainbow-colored 
bar in Fig.~\ref{FIG01} shows the location of the profiles. For an easy inspection, the colors in the 
color bar are associated to the colors of the 44 profiles shown in Fig.~\ref{FIG04}. We do not plot the 
observed profiles to avoid a crowded display. Instead, Fig.~\ref{FIG04} shows the synthetic profiles 
computed from the inferred parameters of the double-Lorentzian fits. One interesting observational 
feature seen in the top panel of Fig.~\ref{FIG04} is that the first profiles (violet profiles) only have 
the slow component. However, once we enter the area of the footpoints (turquoise-green 
profiles) a clearly fast component appears. The shifted profiles next to the pore 
exhibit the strongest redshifts (green-yellow profiles) but possess the smallest amplitudes. The bottom 
panel of Fig.~\ref{FIG04} shows the individual Lorentzian profiles, which yield the dual-flow profile 
shown in the top panel. Blue colors always show the slow component while red colors show the fast 
component. The slow component is always stationary around the reference laboratory wavelength 
10830.30~\AA, whereas the fast component is always redshifted reaching values of up to 10831.40~\AA\ in 
the line core.

Random and systematic errors can influence the measurements. Using the noise estimate 
($\sigma = 3.3$\%) based on continuum intensity variations (see Sect. 2.1) and the fit parameters from 
Eqns.~1 and 2 for single Lorentzian profiles, mock spectra are created and analyzed in the same way as 
the observed data. The linear correlation coefficient for amplitude and HWHM is $\rho=0.99$ and that for 
the velocity is $\rho=0.94$. However, higher Doppler speeds are underestimated. The 
3$\sigma$-uncertainties for amplitude, velocity, and HWHM are 2.2\%, 210~m~s$^{-1}$, and 1.17~pm, 
respectively, whereas the mean values and the shape of the distributions are barely affected. These 
random errors also apply to the double-Lorentzian fits, which are however additionally influenced by 
systematic errors. Dual-flow profiles imply high velocities for the fast component. Therefore, the slow 
component of the \ion{He}{i} red component will be blended with the fast component of the 
\ion{He}{i} blue component.

Using the fit parameters of Eqns.~3 and 4 for the red \ion{He}{i} doublet and scaling 
them appropriately for the blue line of the \ion{He}{i} triplet, mock spectra are generated to obtain an 
error estimate. The aforementioned blend mostly impacts the blue wing of the slow component of the 
\ion{He}{i} red component, which leads on average to a $-300$~m~s$^{-1}$ zero point offset for 
the velocity and a  3$\sigma$-uncertainty of 1.24~km~s$^{-1}$. On average the amplitude of the slow 
component is overestimated by 8\% with a  3$\sigma$-uncertainty of about 27\%. In principle, these 
systematic errors can be reduced either by fitting the entire \ion{He}{i} triplet or by appropriately 
modeling the \ion{He}{i} triplet's blue line when fitting the red doublet. In summary, random and 
systematic errors are sufficiently small to allow a scientific interpretation of the 2D maps of physical 
parameters (for example those in Fig.~2) and of dual-flow \ion{He}{i} spectral profiles.

%##############################################################################
%
%    CONCLUSIONS AND OUTLOOK
%
%##############################################################################

\section{Conclusions and outlook}\label{SEC5}

In this paper we introduce a simple and rapid technique for determining the velocity of multiple 
atmospheric components within a single spatial pixel. This technique produces good fits and gives 
reasonable velocity values, suggesting that after some further tests it can be applied to 
additional data. For example, an implicit assumption is that the \ion{He}{i} line is always optically 
thin in the wavelength range of the telluric line and has to be tested. 

We presented GRIS observations of an EFR containing two small pores. The observations were complemented 
with simultaneous imaging spectropolarimetry acquired with the GFPI in the photospheric \ion{Fe}{i} 
6302~\AA\ line. The FOV for the broad-band images is shown in Fig.~\ref{FIG05} (red box). In addition, 
simultaneous observations with CRISP were carried out. Several full spectropolarimetric scans of the 
photospheric \ion{Fe}{i} 6173~\AA\ line and the chromospheric \ion{Ca}{ii} 8542~\AA\ line were taken 
with a FOV of 54$\arcsec\times$54$\arcsec$ (blue box in Fig.~\ref{FIG05}).

The remaining task will be to extract all the information of the available data sets to understand 
the physical processes that lead to the formation, evolution, and disappearance of these filamentary 
structures. The photospheric \ion{Fe}{i} 6302~\AA\ and 6173~\AA\ lines from GFPI and CRISP will be 
used to study the dynamics and temporal evolution of photospheric structures. 
In addition, the former data will be complemented by the photospheric lines acquired with GRIS, 
e.g., \ion{Si}{i} 10827~\AA, \ion{Ca}{i} 10834~\AA, and \ion{Ca}{i} 10839~\AA, in order to track the
dynamics along several heights within the photosphere. The dynamics in the chromosphere will be covered by the  
\ion{He}{i} 10830~\AA\ triplet as well as with the \ion{Ca}{ii} 8542~\AA\ line. 
By analyzing all available GRIS scans we will be able to further characterize the dual-flow profiles seen in the 
\ion{He}{i} triplet and thus elucidate the origin of the two populations shown in Fig.~\ref{FIG03}.
The link between the chromospheric filamentary structures and the underlying photosphere is still an open question. 
Does the plasma of the chromospheric structures, which exhibit supersonic LOS downflows near the footpoints, 
reach the photosphere? Are the loops also seen in the corona and if yes, what is their behavior? SDO can provide 
such information (e.g., in the quiet corona and upper transition region with
\ion{Fe}{ix} 171~\AA\ EUV images) to study the evolution of the loops in the corona. The magnetic field 
information of the photosphere and the chromosphere will be obtained based on high-resolution imaging 
spectropolarimetric data observed with CRISP. 

Regarding the method to fit the dual flows, we need to expand our study to different data sets. The goal 
is to build a database, which includes many different types of \ion{He}{i} 10830~\AA\ profiles to 
automatically detect the ones with dual-flow components. This database will include profiles extracted 
from (more mature) active regions, sunspots, arch filament systems, and quiet-Sun 
regions. The $\chi^2$-statistics of the fit can provide hints, where a single 
component fit is not sufficient. However, noise and the strength of the \ion{He}{i} absorption 
feature potentially obfuscate such a simplistic approach. Therefore, an automatic detection of dual-flow 
profiles has to rely on easily measured line properties as, for example, the line asymmetry. Additional 
improvements of the algorithm concern the blue component of the \ion{He}{i} triplet. In double-peaked 
profiles, the fast blue component blends with the slow red component, thus affecting the accuracy of the 
double-Lorentzian fitting method.

%##############################################################################
%
%    ACKNOWLEDGEMENTS
%
%##############################################################################

\acknowledgements
The 1.5-meter GREGOR solar telescope was built by a German consortium under the leadership of the 
Kiepenheuer-Institut f\"ur Sonnenphysik in Freiburg with the Leibniz-Institut f\"ur Astrophysik Potsdam, 
the Institut f\"ur Astrophysik G\"ottingen, and the Max-Planck-Institut f\"ur Sonnensystemforschung in 
G\"ottingen as partners, and with contributions by the Instituto de Astrof\'{\i}sica de Canarias and the 
Astronomical Institute of the Academy of Sciences of the Czech Republic. SDO HMI data are 
provided by the Joint Science Operations Center -- Science Data Processing. SJGM is grateful for 
financial support from the Leibniz Graduate School for Quantitative Spectroscopy in Astrophysics, a joint 
project of AIP and the Institute of Physics and Astronomy of the University of Potsdam. PG acknowledges 
the support from grant VEGA 2/0004/16. CD has been supported by grant DE 
787/3-1 of the German Science Fundation (DFG). MC aknowledges the support by the Spanish Ministry of 
Economy and Competitiveness through the project AYA2010-18029 (Solar Magnetism and Astrophysical 
Spectropolarimetry) for the development of the instrument GRIS. This study is supported by the European 
Commission's FP7 Capacities Programme under the Grant Agreement number 312495.

 %###############################################################################
%#
%#    BIBLIOGRAPHY
%#
%###############################################################################

%\input{bibliography.tex}
\bibliographystyle{an}
\bibliography{an-jour,sjgm}

\end{document}